\documentclass[letterpaper, 10 pt, conference]{ieeeconf}
\usepackage[utf8]{inputenc}
\usepackage{amsmath, txfonts, cmll, stmaryrd, authblk}
\usepackage{biblatex, syntax, graphicx}
\usepackage[ruled,vlined]{algorithm2e}

\usepackage{tikz}
\usetikzlibrary{arrows}

\IEEEoverridecommandlockouts      
\makeatletter
\def\BState{\State\hskip-\ALG@thistlm}
\makeatother

\DeclareMathOperator{\PSh}{\mathbf{PSh}}

\DeclareMathOperator{\At}{At}

\DeclareMathOperator{\Unit}{Unit}
\DeclareMathOperator{\force}{force}
\DeclareMathOperator{\lift}{lift}
\DeclareMathOperator{\Let}{let}
\DeclareMathOperator{\In}{in}

\title{Hybrid dynamical type theories for navigation \thanks{This work has been submitted to the IEEE for possible publication. Copyright may be transferred without notice, after which this version may no longer be accessible.}}
\author[1]{Paul Gustafson\thanks{pgu@seas.upenn.edu}}
\author[2]{Jared Culbertson\thanks{jared.culbertson@afresearchlab.com}}
\author[1]{Daniel Koditschek\thanks{kod@seas.upenn.edu}}
\affil[1]{School of Engineering and Applied Science, University of Pennsylvania,
Philadelphia, PA 19104, USA}
\affil[2]{Autonomous Capabilities Team (ACT3), Air Force Research Laboratory, Dayton, OH 45433, USA}

\date{\today}

\begin{document}

\maketitle

\begin{abstract}
We present a hybrid dynamical type theory equipped with useful primitives for organizing and proving safety of navigational control algorithms.    This type theory combines the framework of Fu--Kishida--Selinger for constructing linear dependent type theories from state-parameter fibrations with previous work on categories of hybrid systems under sequential composition.  We also define a conjectural embedding of a fragment of linear-time temporal logic within our type theory, with the goal of obtaining interoperability with existing state-of-the-art tools for automatic controller synthesis from formal task specifications.   As a case study, we use the type theory to organize and prove safety properties for an obstacle-avoiding navigation algorithm of Arslan--Koditschek as implemented by Vasilopoulos.  Finally, we speculate on extensions of the type theory to deal with conjugacies between model and physical spaces, as well as hierarchical template-anchor relationships.
\end{abstract}

\section{Introduction}
The problem of composing complex systems from simpler subcomponents is common to all branches of engineering.   A general approach is to specify interfaces between components -- both logical and physical.  For logical interfaces, arguably the most advanced specification language is dependent type theory, a formal system capable of expressing the entirety of constructive mathematics \cite{martin1982constructive}.   On the other hand, physical interfaces must deal with finite resources, leading naturally to linear logic, a simultaneous refinement of intuitionistic and classical logic that emphasizes the role of formulas as resources \cite{girard1987linear}.    Combining linear logic with dependent types is still an active research area, but much progress has been made in recent years \cite{cervesato2002linear, vakar2014syntax, krishnaswami2015integrating, mcbride2016got, licata2017fibrational}.  We are particularly interested in the categorical framework of \cite{fu2020linear}, in which the nonlinear dependent types can only depend on the ``shape'' of a linear term.  Roughly speaking, this means that type dependency lives on the parameter level and cannot inspect state-level information of linear terms. 

For applications to robotics, we desire a type system capable of encoding sequential and parallel compositions, as well as safety constraints.  The former two forms of composition fit naturally within the framework of linear logic, particularly via its incarnation as the internal language of symmetric monoidal categories. Indeed, a weak version of such a category of directed hybrid systems compatible with sequential and independent parallel composition has been defined in \cite{culbertson2020formal}.  On the other hand, safety constraints have a natural formulation in the internal language of presheaves over an agent's sensorium, a setting particularly well-suited to formulating local conditions in the presence of uncertainty.

As we note below, both of these categories---directed hybrid systems and presheaves over the sensorium---can be connected via a state-parameter fibration, the primary input to the linear dependent type theory construction of \cite{fu2020linear}.   Although that paper  focuses primarily on a symmetric monoidal category of quantum circuits and a corresponding state-parameter fibration from a formal completion of the circuit category to the category of sets, their construction applies equally well to state-parameter fibrations from other symmetric monoidal categories to other locally cartesian closed categories.  In our setting, we replace quantum circuits with directed hybrid systems and the base category $\mathbf{Set}$ with  $\PSh(\Sigma)$, the category of presheaves over a  sensorium $\Sigma$. 

With a state-parameter fibration and its corresponding linear dependent type in hand, we are ready to give types to hybrid controllers used by real engineers. We first provide a conjectural translation from a fragment of linear-time temporal logic (LTL) into the type theory with the goal of interfacing with existing state-of-the-art methods for controller synthesis from formal task specifications \cite{kress2009temporal, saha2014automated}. Then as a case study, we type various components of a navigational controller designed by Arslan--Koditschek \cite{arslan2019sensor} and implemented by Vasilopolous \cite{vasilopoulos2020reactive}. 

\section{Related work}

The closest related work within robotics is the body of literature on controller synthesis from temporal logic specifications. In 1977, Amir Pnueli first proposed linear-time temporal logic as a specification language for formal verification of computer programs \cite{pnueli1977temporal}.   In 1996, Antoniotti and Mishra used computation tree logic, a branching-time temporal logic developed by Clarke and Emerson to analyze concurrent programs \cite{emerson1982using}, to generate a supervisory controller for a walking robot \cite{antoniotti1995discrete}.  In the mid 2000s, Kress-Gazit, Fainekos, and Pappas developed a procedure for automatic controller synthesis from an robot model, class of admissible environments, and LTL task specification \cite{kress2009temporal}.  Subsequently as the field of robotics has grown in importance, temporal logic-based specification has remained the dominant paradigm for formal verification of controllers  \cite{luckcuck2019formal}.

The main difference between our type-theoretic formalization and temporal logic-based approaches is ease of composition.  Whereas temporal logics are close to natural human language \cite{kress2008translating}, their basic connectives (e.g., ``and,''  ``or,'' and classical implication) are a less convenient setting for sequential and parallel composition of dynamical-system-based controllers than the linear logic corresponding to the symmetric monoidal category of directed systems \cite{culbertson2020formal}.  

On the other hand, the relative simplicity of LTL allows for automatic controller synthesis once one restricts to a computationally tractable class of formulas \cite{bloem2012synthesis}.   Thus, the strengths of the type-theoretic and temporal logic approaches are in fact complementary. We provide a conjectural translation between the two logical systems below.

On the type theory side, many have worked on integrating linear logic with non-linear intuitionistic logic in general, and dependent type theory in particular.  The idea of using a linear-nonlinear adjunction, i.e. a pair of symmetric monoidal functors between a symmetric monoidal closed category and a cartesian closed category, to relate a linear type theory to the simply typed lambda calculus originates with Benton \cite{benton1994mixed}.  Adding dependent types is significantly harder since it is not immediately clear what it means for a type to depend on a linear term.  In particular, if $u : L$ is a linear term and $T_x$ is a type family depending on $L$, then both terms $x_u$ of type $T_u$ and the type $T_u$ itself generally reference the term $u$---a violation of linearity, the axiom that linear terms must be used exactly once.  

One approach to resolving this contradiction is to have two separate contexts---one for linear data, and one for intuitionistic data---and prescribe how types are allowed to depend on linear terms.  This is the approach taken by Cervesato and Pfenning in their Linear Logical Framework combining linear logic with $LF$ \cite{cervesato2002linear}.  In their system,  types may not depend on linear terms at all, only intuitionistic terms.

A significant further advance was McBride's use of indices $k \in \{0,1,\omega\}$ to decorate the typing annotations $x :_k A$ to denote the number of times the variable $x$ is used.  When $k = \omega$, his dependent types corresponds to  those of Cervesato and Pfenning.  The index $k=0$ corresponds to terms that occur in types, and $k=1$ corresponds to terms evaluated at runtime.  This allowed much more granular control over variable usage, enabling non-trivial type dependence on linear terms.

Recently, Licata, Shulman, and Riley vastly generalized this work in their fibrational framework for substructural and modal logics \cite{licata2017fibrational}, parameterizing a type theory by an underlying mode theory.  Categorically, their framework corresponds to a functor between 2-dimensional cartesian multicategories, a powerful but very abstract setting.

In the current paper, we employ the framework of Fu--Kishida--Selinger \cite{fu2020linear} for constructing linear dependent type theories from state-parameter fibrations.  Their approach uses McBride's approach to variable count indexing, but also shares the fibrational approach to categorical semantics of Licata--Shulman--Riley in a less abstract setting (monoidal category theory, rather than higher category theory).

\section{A state-parameter fibration}

 The primary datum required in the construction of the linear dependent type theory defined in \cite{fu2020linear} is a state-parameter fibration, a categorical fibration from a symmetric monoidal closed category (a setting for linear logic) to a locally cartesian closed category (a setting for dependent type theory) satisfying some additional compatibility axioms.

To construct such a fibration, we start with a symmetric monoidal category of ``generalized circuits,'' in our case directed hybrid systems as defined in \cite{culbertson2020formal}.    In brief (using the notation of \cite{culbertson2020formal}, where full details can also be found), a hybrid system $H$ consists of a directed graph $G(H)$, whose vertices index a set of continuous modes $\{I_v\}_{v \in G(H)}$ and whose edges index a set of reset maps $\{r_e\}_{e \in G(H)}$ between these modes. We can then define a hybrid semiconjugacy $\alpha: H \to K$ of hybrid systems to be a graph morphism on the underlying graphs together with a collection of smooth maps $\alpha_v: I_v(H) \to I_{\alpha(v)}(K)$ restricting to classical smooth semiconjugacies of the continuous flows and respecting the discrete jumps in the sense that the squares
\begin{equation*}
    \begin{tikzpicture}[baseline=(current bounding  box.center)]
        \node (x) at (0,0) {$I_v(H)$};
        \node (y) at (4,0) {$I_u(H)$};
        \node (z) at (0,-2) {$I_{\alpha(v)}(K)$};  
        \node (w) at (4, -2) {$I_{\alpha(u)}(K)$};
        
        \draw[->, above] (x) to node {$r_e$} (y);
        \draw[->, left] (x) to node {$\alpha_v$} (z);
        \draw[->, right] (y) to node {$\alpha_{u}$} (w);
        \draw[->, below] (z) to node {$r_{\alpha(e)}$} (w);
    \end{tikzpicture}
    \end{equation*}
commute for each reset map $r_e: I_v \to I_u$. These basic constructions then allow us to define a directed hybrid system to be a cospan of hybrid semiconjugacies
\[
    H_i \to H \leftarrow H_f
\]
such that (i) both legs are embeddings, (ii) each component map of the right leg is a diffeomorphism, (iii) the image of the graph $G(H_f)$ in $G(H)$ is a sink, and (iv) for every $\epsilon, T > 0$ and state $x \in I(H) = \sqcup_{v \in G(H)}I_v(H)$, there exists an $(\epsilon, T)$-chain from $x$ to the image of $H_f$ in $H$.   Roughly speaking, such a system forms the ambient substrate for a ``funnel'' leading its initial subsystem $H_i$ into its final subsystem $H_f$.

Sequential composition corresponds to the usual notion of composition of cospans via pushouts: the directed system $H' \odot H$ obtained by composing the systems $H: H_i \to K$ and $H': K \to H'_f$ is given by 
\[
    \begin{tikzpicture}
        \node (x) at (0,0) {$K$};
        \node (y) at (3,0) {$H'$};
        \node (z) at (0,-2) {$H$};
        \node (w) at (3, -2) {$H' \odot H$\,,};

        \draw +(2,-1.5) -- +(2,-1)  -- +(2.5,-1);

        \draw[->, above] (x) to node {$K \to H'$} (y);
        \draw[->, left] (x) to node {$K \to H$} (z);
        \draw[->, right] (y) to node {} (w);
        \draw[->, below] (z) to node {} (w);
    \end{tikzpicture}
\]
an operation defined only up to isomorphism \cite{benabou1967introduction,grandis1999limits}.  Thus, in order to get a strict category,  we will consider directed systems up to conjugacy; that is, up to isomorphisms with respect to hybrid semiconjugacy.   Using the categorical product with respect to semiconjugacy defined in \cite{culbertson2020formal} as the monoidal product, we then have a symmetric monoidal category $\mathbf{DH}$ of directed systems up to conjugacy under sequential composition.

Following the recipe of \cite{fu2020linear}, our first step towards constructing a corresponding linear dependent type theory is to embed $\mathbf{DH}$ into a symmetric monoidal closed category with products $\mathbf{\overline{DH}}$ by taking the Yoneda embedding.   This procedure formally extends the category with exponential objects and categorical products for a more ergonomic type theory.  

The second step is to form a category $\mathbf{\overline{\overline{DH}}}$ of parametrized circuits.  In their case, the spaces of parameters are sets; in ours, presheaves over a topological space $\Sigma$, which we call the sensorium. An object $A$ of this category consists of a pair $(\underline{A}, (A_x)_{x \in A})$ where $\underline{A}$ is a presheaf over the sensorium $\Sigma$, and $(A_x)$ is an indexed family of objects of $\mathbf{\overline{DH}}$, one for each element of $A$ (here, an element of a presheaf is a morphism $* \to A$, i.e. a choice of set-element over every open set of $\Sigma$).  An arrow $f : A \to B$ consists of a pair $(\underline{f}, (f_x)_{x \in A})$ where $\underline{f} : \underline{A} \to \underline{B}$ is a map of presheaves, and $(f_x : A_x \to B_{\underline{f}(x)})$ is an indexed family of morphisms of $\mathbf{\overline{DH}}$.  

One can check, following the proof of Fact~2.4 in \cite{fu2020linear}, that the functor $\mathbf{\overline{\overline{DH}}} \to \PSh(\Sigma)$ given by projecting to the first component forms a state-parameter fibration.

\section{Type system}
For the purposes of our navigational case study, we focus on three families of simple types:  $See(n)_{n \in \mathbb{N}}$, $At(x)_{x \in X}$ where $X$ is a two-dimensional manifold (the ``workspace''), and $Safe(c)$ where $c$ is any term.  Figure~\ref{tab:semantics} provides their semantics as objects in $\mathbf{\overline{\overline{DH}}}$, fixing the sensorium $\Sigma := C(S^1, \overline{\mathbb{R}_{\ge 0}})$ corresponding to distance measurements from a limited range line-of-sight sensor.  For example, the hybrid system associated to $See(n)$ consists of $n$ object centers and radii with a $0$ vector field, while its presheaf consists of a proof that for every sensor measurement $f$ in an open set of uncertainty $U$, there exist $n$ connected components of distance readings less than a maximum range $M$.  

\begin{figure}
\begin{tabular}{ c c c }
 Type & Hybrid System & Presheaf (evaluated at $U \subset \Sigma$)  \\ 
 \hline
 $See(n)$ & $(X^n \times \mathbb{R}^n, 0)$ & $|\pi_0(f^{-1}([0,M])| = n \quad \forall f \in U$  \\  
 $At(g)$ & $(X, -\nabla\|x - g\|^2)$ & $\sup_{f \in U} d(x_f,g_f) < \epsilon$  \\ 
 $Safe(c)$ & $Unit$ & $\sup_{f \in U} \min_\theta f(\theta) > R$

\end{tabular}
\caption{Semantics of simple types with sensorium $\Sigma := C(S^1, \overline{\mathbb{R}_{\ge 0}}$).  In the presheaf associated to $\At(g)$, the symbols $x_f$ and $g_f$ correspond to the robot's inferred position and goal locations based on a sensor reading $f$. }
\label{tab:semantics}
\end{figure}

With the exception of the simple types, the syntax of the corresponding type theory shown in Figure~\ref{fig:syntax} is identical to \cite{fu2020linear}, which we refer to for the corresponding inference rules.  The most interesting types are the linear dependent sum and linear dependent function.   Just as with regular dependent types, a term of a linear dependent sum $(x:A) \otimes B[x]$ is a pair $(x, b_x)$ of types $A$ and $B[x]$, respectively, and a term of a linear dependent function $(x:A) \multimap B[x]$ is a function that returns a term of type $B[x]$ for every term $x:A$.  However, linear (i.e. state-level) variables can only be used at most once with one exception---their ``shape'' can be used arbitrarily many times in parameter types.  Roughly speaking, the shape of state-level data corresponds to replacing every simple type in its definition by the $Unit$ parameter type.

In the formal type theory, variable usage is encoded by the index $k$ in $x :_k A$, and the inference rules enforce state-parameter usage constraints.  The ``!'', $\lift$, and $\force(')$ operations move types and terms from the state-level to the parameter level and vice versa, while ensuring that parameters can only depend on the shape of linear terms.   

\begin{figure}
\begin{itemize}
  \item[] \emph{Simple types} $S ::= See(n) \mid  At(x) \mid Safe(M)$
  \item[] \emph{Types} $A,B ::= S \mid Unit \mid !A \mid (x:A) \multimap B[x] \mid (x:A) \otimes B[x] \mid (x:P_1) \to P_2[x]$
  \item[] \emph{Parameter types} $P ::= Unit \mid !A \mid (x : P_1) \otimes P_2[x] \mid (x : P_1) \to P_2[x]$
  \item[] \emph{Terms} $M,N,L ::= unit \mid x \mid \lambda x.M \mid M N \mid \force M \mid \force' R$ \\
  $\mid \lift M \mid (M, N) \mid \Let (x,y) = N \In M \mid \lambda'x.R \mid R_1 @ R_2 $
  \item[] \emph{Parameter terms} $R ::= unit \mid x \mid \lambda' x.R \mid R_1 @ R_2  \mid \force' R$ \\
  $\mid \lift M \mid (R_1, R_2) \mid \Let (x,y) = R_1 \In R_2 $
  \item[] \emph{Values} $V ::= unit \mid x \mid \lambda x. M \mid \lambda' x. R \mid \lift M$
  \item[] \emph{Indices} $k ::= 0 \mid 1 \mid \omega$
  \item[] \emph{Contexts} $\Gamma ::= \cdot \mid x :_k A , \Gamma$
  \item[] \emph{Parameter contexts} $\Phi ::= \cdot \mid x :_k A, \Phi$, where $k = 1$ only if $A$ is a parameter type.
\end{itemize}
\caption{Syntax of navigational type theory}
\label{fig:syntax}
\end{figure}

We also have a conjectural embedding of a fragment of linear temporal logic (LTL) \cite{pnueli1977temporal} into the type theory shown in Figure~\ref{tab:ltl}.  We translate the atomic propositions directly into hybrid dynamical types whose presheaves enforce the invariant; for example, inhabitation in a partition element corresponds to a presheaf only supported on that element.  We translate ``or'' as the disjoint union of hybrid systems, and ``and'' as the parallel composition.  The ``Eventually'' diamond corresponds to a directed system with codomain corresponding to the argument of the diamond operator.  We simply translate the ``Always'' box operator as a directed system whose corresponding presheaf always validates the argument proposition.

\begin{figure}
\begin{tabular}{ c c  }
 LTL Proposition $p$ &  Hybrid dynamical type $T(p)$  \\
 \hline
 $p \vee q$ & $T(p) \oplus T(q)$   \\
 $p \wedge q$ & $T(p) \otimes T(q)$ \\
 $\Diamond p$ & $\Unit \multimap T(p)$ \\
 $\Box p$ & $T(p)$ 
\end{tabular}
\caption{Conjectural embedding of a fragment of LTL}
\label{tab:ltl}
\end{figure}

\section{Navigation example}

\begin{figure}
    \centering
    \includegraphics[width=0.4\textwidth]{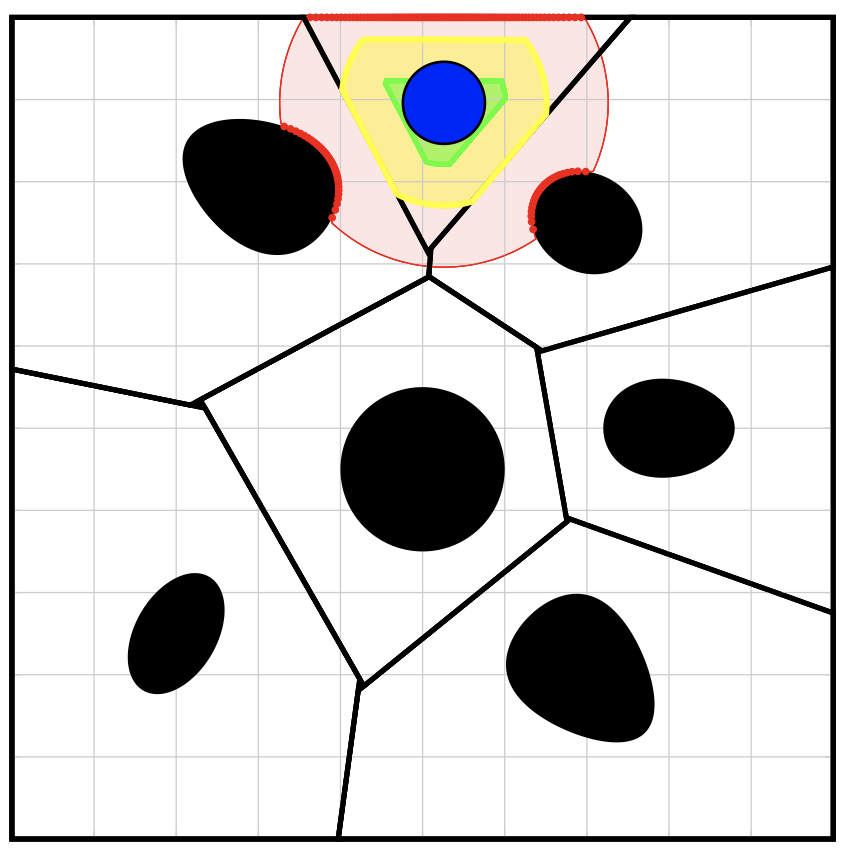}
    \caption{Voronoi-based navigation using a limited range line-of-sight sensor (Image source: \cite{arslan2019sensor})}
    \label{fig:voronoi}
\end{figure}

As a case study, we formulate the navigation algorithm of \cite{arslan2019sensor} as a composition of typed hybrid systems and proofs of correctness. Under an assumption on the curvature of the obstacle boundaries (satisfied by circles, for example), the main theorem of the paper is essentially
$$(d(O_i, O_j) > 2R \, \wedge \, d(x_0, O_i) > 2R) \to \Box \forall_i d(O_i) > R \, \wedge\,  \Diamond(At(g)),$$
where $x_0$ is the initial position of the robot, $R$ is a safety radius. 
Rearranging things slightly, we want
\begin{align*}
d(x_0, O_i) > 2R) & \to \exists_{i,j} (d(O_i, O_j) \ge 2R \\
& \vee \,(\Diamond(At(g))  \, \wedge\,  \Box \forall_i d(O_i) > R).
\end{align*}

Under the translation shown in Figure~\ref{tab:ltl}, our task is to construct a controller of type 
\begin{align*}
(d(x_0, &  O_i) > R) \to (O_i, O_j : Obstacle) \otimes (d(O_i, O_j) \ge 2R) \\
& \oplus (c : (Unit \multimap At(g) \oplus  \, d(O_i, O_j)  \leq 2R)) \otimes Safe(c).
\end{align*}
That is, if the initial position of the robot is safe, then we have a safe controller that reaches the goal unless the obstacle-separation assumption is violated.

\begin{figure*}
 \centering
\begin{align*}
Obstacle & = (center : X) \times (radius : \mathbb{R}_{\ge 0}) \\
SeparationViolation & = (O_i, O_j : Obstacle) \to d(O_i, O_j) \leq 2R \\
go      & :  (g : X, n: \mathbb{N}) \to (c : (s : See(n)) \multimap (At(g) \otimes See(n)) \oplus Interrupt(s))\\
Interrupt & : See(n) \multimap NewObs(See(n+1))
 \oplus LoseObs(See(n-1))  \\
 & \oplus SV((O_1, O_2: Obstacle) \otimes SeparationViolation (o1, o2)) \\
detect & : See(n) \multimap See(n-1) \oplus See(n) \oplus  See(n+1) \\
visibleObs & : \,!See(n) \to List(Obstacle) \\
projGoal & : ConvHull(n) \to X \to X \\
voronoi & : \,!See(n) \to ConvHull \\
ConvHull & = List(X) \\
startSensing & : Unit \multimap (n : \mathbb{N}) \otimes  See(n)\\ 
stopSensing & : (n: \mathbb{N}) \to See(n) \multimap Unit \\ 
controller & : (g : X) \to d(x, nearestObs(s)) > R)  \\
& \to (f: Free \multimap At(g) \oplus (O_i, O_j : Obstacle) \otimes SeparationViolation(O_i, O_j)) \otimes Safe(f)
\end{align*}
\caption{Types for navigation functions}
\label{fig:types}
\end{figure*}

\begin{figure*}
\begin{verbatim}
controller g p = (f, safetyProof f ) where
  f = stopSensing . c . startSensing
  c s = case mpg g p s of  
    Left (localGoal, s)) -> 
      if localGoal == g
      then (sp, d)
      else mpg g f s
    Right (NewObs s')  -> inl (mpg g f s')
    Right (LoseObs s') -> inl (mpg g f s')
    Right (SeparationViolation o1 o2) -> inr (SeparationViolation o1 o2)
  mpg g p s = go(projGoal(voronoi(lift(s)), g), f, s))
  safetyProof f = <Euclidean geometry formalization of Arslan-Koditschek proof>
\end{verbatim}
\caption{Navigation controller pseudocode}
\label{fig:controller}
\end{figure*}

The types shown in Figure~\ref{fig:types} provide some Haskell-style typing declarations for the functions involved in such a controller.
For example, the ``go'' function's typing declaration in Figure~\ref{fig:types} has the following meaning---for every natural number $n$ and goal location $g$, given the data of a $Free$ navigational state and $n$ visible objects, this controller will either end up at the goal while still seeing $n$ objects or it will be interrupted.  The $Interrupt$ type says that the only interrupts correspond to seeing a new object, forgetting an object, or noting that the object separation axiom is violated.

The ``controller'' function (Figure~\ref{fig:controller}) is defined by looping over the MPG (move-to-projected-goal) function as follows.   The result of one application of the MPG function is either arriving at the projected goal or an interrupt.   If you have arrived at the projected goal, check to see if it is the final goal point, in which case you are done.   Otherwise, check to see if you have detected an obstacle separation violation.  If so, pass the violation along and stop. In all other cases, perform MPG again with your current location and sensor readings. In addition, one would like a proof of safety --- i.e. a term of the $Safe$ type taking the controller definition as an argument.   This corresponds to formalizing Proposition~3 of \cite{arslan2019sensor}, which states that the robot's free space is positively invariant.


\section{Future work}

This paper has outlined a type theory useful for organizing and proving safety properties for navigational tasks -- a baby step towards our broader goal of a functional programming language for physical work.  On the practical side, the next step is to integrate some form of type checking with physical robot platforms.   There are two parts to this task -- (i) integrating the Robot Operating System (ROS) \cite{quigley2009ros} with better support for functional programming in general and linear dependent types in particular, and (ii) completing the formalization of the algorithm in Figure~\ref{fig:controller}.  For the first task, one approach would be to use ideas from RosHask \cite{cowley2011stream} to call ROS from Haskell \cite{jones2003haskell}, and either embed a linear dependent language in Haskell or follow Agda's approach to foreign function interface to Haskell \cite{norell2008dependently}.  For the second task, the main obstacle is the reliance of the proof of safety on basic Euclidean geometry which requires some effort to formalize.  Fortunately, it may be possible to leverage GeoCoq \cite{boutry:hal-01483457}, a dependently typed formalization Tarski's axiomatic system \cite{tarski1999tarski} in the Coq proof assistant \cite{coquand1986calculus, barras1997coq}, previously used to formalize the first book of Euclid's Elements \cite{beeson:hal-01612807, heath1956thirteen}.

More theoretically, the type theory still leaves much to be desired.  Perhaps most urgent is the inclusion of template-anchor relationships, a form of controller hierarchy integral to much of robotic programming \cite{full1999templates}.  A first step towards this goal would be to extend the type theory to deal with hybrid semiconjugacies, the vertical morphisms in the double category of hybrid systems defined in \cite{culbertson2020formal}, potentially leveraging work of New and Licata on double categorical semantics of gradual typing \cite{licata2020call}.  

Another important direction is the exploration of more expressive categories of ``funnel-like'' systems closed under sequential composition.  For example, since our category of directed systems is not monoidal closed, we need to use the Yoneda embedding to formally adjoin exponential objects.  These formal exponentials are not hybrid systems, but rather spaces of recipes for producing hybrid systems.  However, just as it is often useful to use the fact that the collection of functions between two finite sets is itself a set, or the collection of linear maps between two vector spaces is a vector space, we believe that putting dynamical structures on the space of ``funnels'' between two systems will prove useful.  However, fundamental obstructions exist when considering even classical dynamical systems, and finding a convenient category in this setting requires considering alternative generalizations of manifolds such as Chen spaces, diffeological spaces, or Fr\"{o}licher spaces \cite{baez2011convenient,frolicher1980categories}. Thus, a more natural setting for exploring exponentials in categories of hybrid systems might be to consider more topological perspectives such as locally preordered spaces \cite{krishnan2009convenient}. Additionally, we might imagine a refinement procedure that takes a crude plan (``funnel'') to get from an initial system to a final system, and iteratively improves the plan, incorporating a hierarchical notion of dynamics.

\section{Acknowledgment}

This work was supported in part by ONR N000141612817, a Vannevar Bush Faculty Fellowship held by Koditschek and by UATL 10601110D8Z, a LUCI Fel-lowship held by Culbertson, both granted by the Basic Research Office of the US Undersecretary of Defense for Research and Engineering.


\printbibliography

\end{document}